\documentclass[manuscript]{acmart2}

\AtBeginDocument{%
  }
    
\copyrightyear{2025}
\acmYear{2025}

\begin{document}

\title{Interactive AI Alignment: Specification, Process, and Evaluation Alignment}

\author{Michael Terry}
\email{michaelterry@google.com}
\affiliation{%
  \institution{Google DeepMind}
}

\author{Chinmay Kulkarni}
\affiliation{%
  \institution{Google}
}

\author{Martin Wattenberg}
\authornote{Work conducted while at Google.}
\affiliation{%
  \institution{Harvard University and Google DeepMind}
}

\author{Lucas Dixon}
\affiliation{%
  \institution{Google DeepMind}
}

\author{Meredith Ringel Morris}
\affiliation{%
  \institution{Google DeepMind}
}

\renewcommand{\shortauthors}{Terry et al.}

\begin{abstract}
  Modern AI enables a high-level, declarative form of interaction: Users describe the \textit{intended outcome} they wish an AI to produce, but do not actually create the outcome themselves. In contrast, in traditional user interfaces, users invoke \textit{specific operations} to create the desired outcome. This paper revisits the basic input-output interaction cycle in light of this declarative style of interaction, and connects concepts in AI alignment to define three objectives for \textit{interactive alignment} of AI: specification alignment (aligning on what to do), process alignment (aligning on how to do it), and evaluation alignment (assisting users in verifying and understanding what was produced). Using existing systems as examples, we show how these user-centered views of AI alignment can be used descriptively, prescriptively, and as an evaluative aid.
\end{abstract}

\maketitle

\section{Introduction}

\begin{figure}[h]
    \centering
    \includegraphics[width=0.65\textwidth]{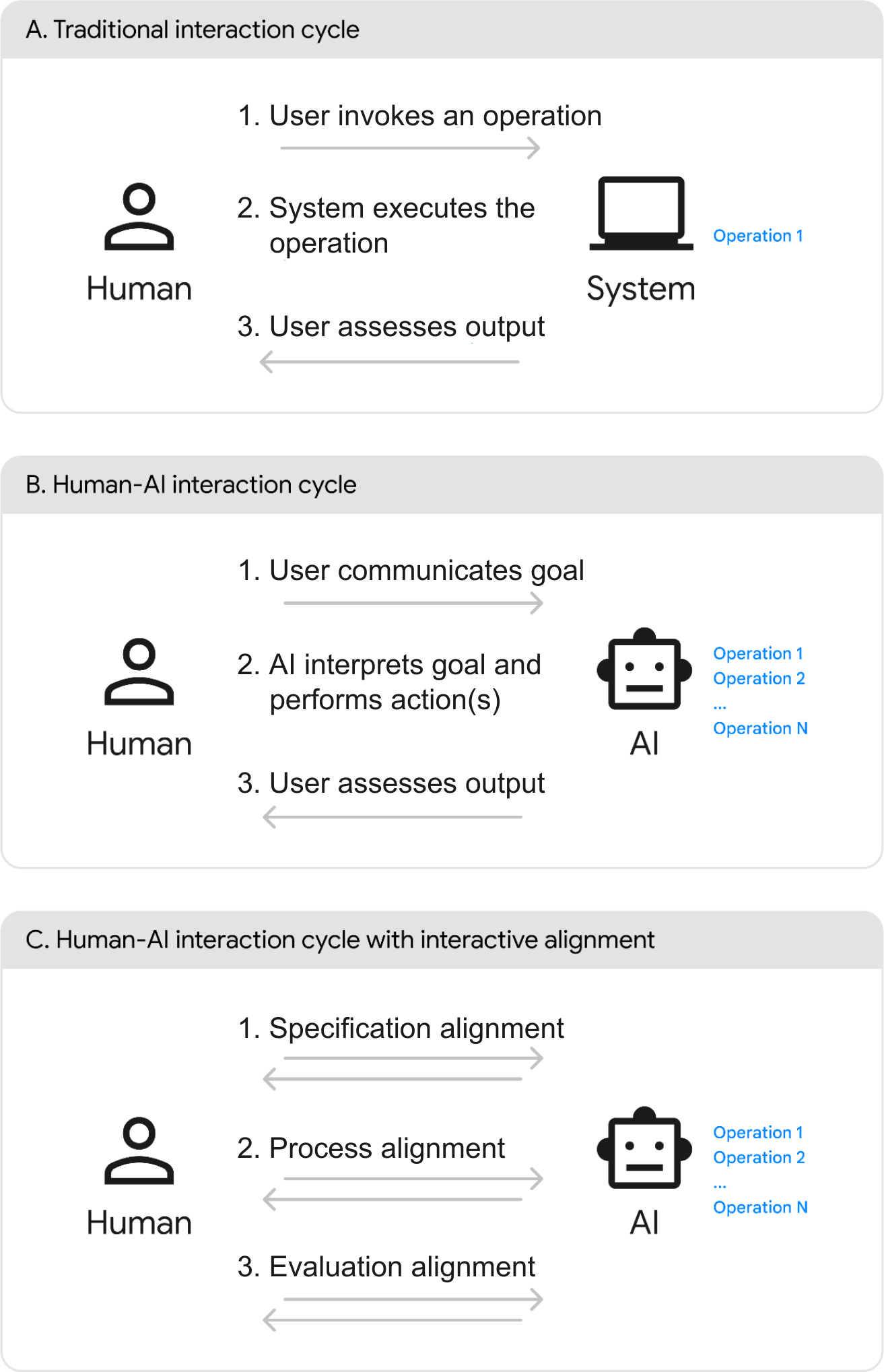}
    \caption{Basic models of interaction. \textbf{A:} In interacting with a traditional non-AI system, the user chooses an operation to perform and provides \textbf{input} to the system to perform that operation (1). The system performs the operation (2), then provides the \textbf{output} to the user, which they assess (3) with respect to their goals. \textbf{B:} When interacting with an AI, the user communicates their \textit{desired outcome} to the AI (1), the AI interprets the goal and performs operations to achieve that goal (2), and (3) the \textbf{output} is sent to the user. \textbf{C:} The same human-AI interaction cycle with AI alignment concepts mapped onto the three steps: (1) \textbf{Specification alignment} mechanisms provide means for the user to align the AI on the specific task to perform. (2) \textbf{Process alignment} mechanisms enable the user to modify how the task is performed, potentially offering the user direct control over specific steps. (3) \textbf{Evaluation alignment} mechanisms help the user assess and understand the output. }
    \Description{Figures showing a) users submitting input to a computer and receiving output, b) users submitting a goal to an AI, it processing the input, and providing output, and c) users submitting a goal to an AI, then interacting with the AI to get alignment on the specification, process, and evaluation (assessment) of the end result.}
    \label{fig:interaction_cycles}
\end{figure}

The basic interaction cycle is defined by 1) the user providing input to a computational system (e.g., keyboard input, pointer input), 2) the system acting on that input, and 3) the system producing an output, which the user assesses to determine their next action, if any (Figure \ref{fig:interaction_cycles}A). Within this basic interaction cycle, Norman defined two gulfs a user must bridge to achieve their goals: The Gulf of Execution (the need to determine how to formulate input to the system to achieve the desired goal) and the Gulf of Evaluation (the need to understand the current system state) \cite{norman_2002}. In traditional interactive systems, user input typically leads to invocation of a specific \textit{operation}. For example, user input may invoke an operation to delete a word (e.g., through keyboard input), navigate to a new URL (e.g., via a button press on a mouse), or draw a line in a painting application (e.g., by the mapping of pointer movements to movements of a paint brush).

Recent advances in AI have led to what some consider a qualitatively different form of interaction \cite{nielsen2023}. In this interaction paradigm, users provide high-level, natural language input that describes the desired \textit{outcome}, rather than a specific operation to invoke. As examples, models now exist to transform natural language input into code \cite{copilot}, images \cite{midjourney}, music \cite{agostinelli2023musiclm}, and video \cite{ho2022imagen}. In the best case, the natural language input (and intent) is correctly interpreted and the model produces the desired result (though this is obviously not always the case; e.g., see \cite{Pereira2023a}). Jakob Nielsen describes this interaction paradigm as ``intent-based outcome specification,'' where the user provides a specification of the outcome, and then the AI, not the person, determines how to achieve that outcome \cite{nielsen2023}.

This paper revisits the basic interaction cycle and Norman's Gulfs of Execution and Evaluation in light of this increasingly popular interaction paradigm\footnote{There is a long history of researching and developing more declarative approaches to human-computer interaction. As an example, declarative languages like SQL and HTML describe the desired outcome, but not the specific operations to perform. Recent advances in machine learning, most notably, large language models \cite{brown2020language}, have arguably made this interaction paradigm far more capable and reliable, and thus worthy of renewed attention.}.
The HCI community has a long tradition of connecting and integrating research threads to identify and describe an emerging interaction paradigm (e.g., Shneiderman's description of direct manipulation \cite{shneidermandm} or Horvitz's principles of mixed initiative interfaces \cite{horvitz_1999}). This paper follows in this tradition and connects and integrates common research threads in the AI alignment and HCI literature to identify three AI alignment objectives within the basic human-AI interaction cycle (see also Table \ref{tab:core-dimensions}):

\begin{enumerate}
  \item \textbf{Specification alignment}: Aligning on \textit{what} the AI should do, which may include aligning the \textit{user} about what the AI can actually do (i.e., \textit{bidirectional alignment} \cite{shen2024bidirectionalhumanaialignmentsystematic}).
  \item \textbf{Process alignment}: Aligning on \textit{how} the AI should achieve the desired outcome (i.e., giving the user some degree of control and transparency over the AI's process).
  \item \textbf{Evaluation alignment}: Designing the model and/or the interface to ensure the user can verify and understand the AI's output.
\end{enumerate}

Collectively, we call this process \textit{interactive alignment}. %

Each form of alignment naturally implies a corresponding gulf for the user to bridge. For example, diffusion models create images starting from statistical noise, a very different process from how people typically create images. Controlling this process requires bridging the \textit{Process Gulf}: Giving users the tools and capabilities that translate user intents into forms that directly control or otherwise steer the AI's process. More generally, for each form of alignment, the user must determine \textit{how} to align the AI, and assess whether alignment has been achieved. In some cases, the AI may also need to align the \textit{user} on what is possible (i.e., what it can do)---that is, alignment can be \textit{bidirectional} \cite{shen2024bidirectionalhumanaialignmentsystematic}. Just as a well-designed traditional user interface helps the user bridge the Gulfs of Execution and Evaluation, a well-designed human-AI interface will help the user bridge these alignment gulfs to reach specification, process, and evaluation alignment.

A number of these concepts have been discussed in varying degrees in past AI alignment and HCI literature (e.g., the notion of a ``specification'' is discussed in both the AI alignment \cite{leike2017ai, ortega_2018} and HCI literature \cite{nielsen2023}). %
This paper creates an explicit mapping between related concepts in these separate communities to 1) highlight common ground and 2) identify opportunities to jointly approach the problem of creating useful, usable, and aligned AI.

\begin{table*}
  \caption{User-centered alignment dimensions in the design and evaluation of interactive AI systems.}
  \label{tab:core-dimensions}
  \def\arraystretch{1.5} %
  \begin{tabular}{|p{1.5in}|p{2in}|p{2in}|}
    \toprule
    \textbf{Term}
    &
    \textbf{Definition}
    &
    \textbf{Example}
    \\
    \bottomrule
    \textbf{Specification alignment}
    &
    The process by which the person and the AI align on \textit{what} the desired outcome is.
    \newline
    \newline
    Issues relevant to specification alignment include 1) under-specified, ambiguous, and/or incorrect user input, and 2) potentially aligning the \textit{user} about what the AI is actually capable of (i.e., bidirectional alignment \cite{shen2024bidirectionalhumanaialignmentsystematic}), which may include providing \textit{AI affordances}.
    &
    Given the input ``Write a TODO app,'' a coding assistant provides a specification of what the system it will write: ``A TODO app written in JavaScript and HTML, with the ability for users to add and delete TODO items.''
    \\\hline
    \textbf{Process alignment} 
    &
    The process of aligning on \textit{how} the AI will produce the desired outcome.
    \newline
    \newline
    Process alignment implies the ability for the user to more directly control the AI's process, but also suggests the value in being able to observe, audit, and/or understand the AI's process (i.e., to provide \textit{transparency} into the AI's process).
    &
    A code synthesis assistant describes which libraries and coding styles it will use in writing the TODO app. %
    \\\hline
    \textbf{Evaluation alignment} 
    &
    The process of the user assessing the AI's output to verify it meets their objective (i.e., did the AI produced the desired outcome?).
    \newline
    \newline
    An even stronger version of evaluation alignment is ensuring the user \textit{understands} the model's output (e.g., in cases where the user may not be familiar with the problem space).
    \newline
    \newline
    Depending on the context (e.g., the user and/or problem area), the user may benefit from explicit mechanisms from the user interface and/or the AI to help them reliably assess and/or understand the AI's output.
    &
    The user can determine whether a coding assistant's code actually creates a functional TODO app meeting their requirements. Specific mechanisms could include running the code in a sandboxed environment and the AI producing appropriately commented code.
    \\
  \bottomrule
\end{tabular}
\end{table*}

Using a set of case studies, we show how these user-centered alignment objectives and their corresponding gulfs provide useful analytical lenses to characterize and assess the designs and capabilities of existing interactive AI systems. We also show how the alignment objectives are \textit{prescriptive} and can suggest opportunities for improving an interactive AI's user interface.

The rest of the paper reviews related work in the domains of AI alignment and HCI research, drawing connections between common themes and goals. We then provide a detailed description of each form of user-centered alignment, and show how they can be used to describe a set of existing interactive AI systems. We conclude by discussing how these concepts relate to Norman's original gulfs, and how the concepts can be practically applied and extended to further move towards the overall goal of creating usable, useful, and aligned AI.

\section{Related Work}
We review related work in AI alignment and interactive systems through the lens of a basic three step interaction cycle, defined as follows (Figure \ref{fig:interaction_cycles}A):

\begin{enumerate}
  \item \textbf{User input}: The user provides input to the system.
  \item \textbf{System execution}: The system acts on that input.
  \item \textbf{System output and user assessment}: The system produces an output that the user assesses.
\end{enumerate}

In the last step, the user evaluates the system output to determine whether it produced the expected result, but also to determine what action (if any) to take next. If further actions are required, the cycle repeats.

After briefly reviewing the fundamental concepts of AI alignment, we consider alignment and user needs for each step of the above cycle. 

\subsection{AI Alignment}
The goal of AI alignment can be thought of as ensuring an AI produces intended outcomes without undesirable side effects \cite{leike2018scalable, christiano2018, kenton2021alignment, kirchner2022researching, christian2020alignment, irving2018ai}. This overall concept can be considered from a wide range of perspectives. For example, Gabriel observes that one could attempt to align an AI with ``instructions, intentions, revealed preferences, ideal preferences, interests and values'' \cite{Gabriel_2020}. Similarly, alignment can be considered from the perspective of system creators, end-users, or society at large. One area of particular concern in the alignment literature is that of safety and reducing the likelihood of harms arising through AI \cite{irving2018ai, russell2016research}.

In this paper, we focus on the problem of AI alignment from the perspective of end-users directly interacting with an AI to accomplish a specific goal. We further constrain our scope to analyzing alignment needs at the level of the basic interaction cycle defined above. Given this scoping, we limit our coverage of otherwise critical topics such as AI safety \cite{russell2016research}, specification gaming \cite{krakovna2020}, and malicious and/or deceptive AI \cite{bostrom_2014}. For a more comprehensive survey of the AI alignment literature, see Shen et al.'s review \cite{shen2024bidirectionalhumanaialignmentsystematic}.

\subsection{Alignment and the Interaction Cycle}
When AI is introduced into an interactive system, it can fundamentally alter the basic interaction cycle \cite{yang2020, amershi_2019}. For example, an AI may incorrectly interpret the user's intent, creating a need for interfaces that anticipate potential inference errors, and additional capabilities to minimize the costs of correcting these errors \cite{amershi_2019}.
Accordingly, a number of resources and guides have been developed to support the design of AI-based systems, including Horvitz's Principles of Mixed Initiative Interfaces \cite{horvitz_1999}, Amershi et al.'s Guidelines for Human-AI Interactions \cite{amershi_2019}, and the People + AI Research Guidebook \cite{pairguidebook}.

In the sections below, we draw from this prior work and consider how AI affects each step of the three step interaction cycle, integrating research threads from both the AI alignment and HCI communities in the process. Throughout the discussion, we refer to an example scenario where a user attempts to write a TODO app using an AI coding assistant.

\subsubsection{Alignment and User Input (Step 1)}
In the first step of the interaction cycle, the user provides input to the system. With common AI systems such as chatbots and generative models, a user inputs the \textit{desired outcome} (i.e., their goal, such as a description of an image to generate), as opposed to specifying the \textit{operation} to perform to achieve the final, desired outcome \cite{nielsen2023}. This process of inputting the intended outcome has been described by both the alignment and HCI communities as providing a \textit{specification} to the system \cite{leike2017ai, ortega_2018, nielsen2023}. In the TODO app example, a user might enter the input ``Write me a TODO app.''

Modern AI systems enable people to explicitly communicate specifications through semantically meaningful inputs such as natural language \cite{bommasani2022opportunities, agostinelli2023musiclm, ramesh2022hierarchical, saharia2022photorealistic, ho2022imagen, copilot} or images \cite{stabledoodle}. However, some systems also proactively \textit{infer} intent from context, such as code synthesis assistants like CoPilot \cite{copilot}, or research systems like Horvitz's LookOut 
\cite{horvitz_1999}. We mostly focus on the former use case in this paper (explicit user input), but note that implicitly determining user intent is a rich space in and of itself.

While natural language input seems like it would lead to usable and useful systems, it can be challenging to fully and accurately communicate an intended outcome \cite{leike2018scalable, irving2018ai}. For example, describing all the desired capabilities of a TODO app requires a fair amount of effort and detail. A user's request may also be ambiguous, under-specified, or even incorrect. For example, the user request of ``Write me a TODO app'' is under-specified as it doesn't provide details such as the programming language to write it in, the platform(s) it should run on, the features to include, etc.

Collectively, the challenge of providing a complete, correct specification is referred to as the \textit{specification problem} \cite{leike2018scalable, ortega_2018}, and motivates the need for mechanisms that support users in interactively aligning an AI to their specific goals.

Recent work has attempted to address the specification problem in large language models (LLMs) through non-interactive approaches such as instruction tuning and reinforcement learning through human feedback (RLHF) \cite{ouyang2022training}. %
However, despite these advances, user-centered research has found that it can still be quite challenging for users to achieve desired outcomes when using natural language input \cite{Pereira2023a, Pereira2023b, jiang2022}. In this work, rather than focus on aligning the model, we emphasize opportunities for \textit{interactive alignment}, or the ability for the end-user to align the AI during application use. 

Some of the challenges users face can be partially attributed to a lack of \textit{affordances}, where an affordance is an element of an interface's design intended to communicate what the system is capable of doing \cite{norman_2002}. Modern graphical user interfaces signal what is possible through mechanisms such as graphical icons, visual design, and hierarchical menus. While it is not uncommon for chatbot systems to provide example inputs to teach people what the system is capable of doing, the wide range of capabilities of generative models can make it difficult to communicate clear boundaries about what is and isn't possible. Thus, while Amershi et al.'s guidelines advocate for explicitly providing an indication of what an AI can do, and how well it can do it \cite{amershi_2019}, there are opportunities for further research into concrete approaches for designing AI interfaces to help people understand the capabilities of highly general AI systems such as those employing foundation models \cite{bommasani2022opportunities}.

Inferring the user's intended specification can also be challenging if their input is ambiguous or under-specified. In this light, Horvitz argues for the value of employing \textit{dialogue} with the user to resolve uncertainty in a user's intentions \cite{horvitz_1999}. 

In this paper, we introduce the concept of \textit{specification alignment}; we demonstrate how this concept can support ideating new interfaces and interactions for conveying the affordances of highly general AI systems and helping end-users clarify under-specified goals.  

\subsubsection{Alignment and System Execution (Step 2)}
In the second step of the interaction cycle, the system executes the operation specified by the user. In traditional interactive systems, the specific operations performed are explicitly designed and implemented by the system developers. With machine learning-based systems, the actual computations performed have been learned through examples, which can lead to additional needs for interactive system design\footnote{AI systems may also make calls to APIs, meaning computations can be spread across multiple computational processes (including other models). But the key feature of this interaction paradigm is the AI controlling the overall computational process.}. In particular, a common finding in human-centered AI research is the desire for users to more directly \textit{control} a model's execution in domains ranging from medical assistants \cite{cai2019} to music generation \cite{louie2020}. The AI alignment literature has described this general problem as the ``control problem'' \cite{bostrom_2014}.

Numerous tools and mechanisms have been developed to offer users greater control over AI, including those that enable users to directly influence search mechanisms \cite{cai2019}, music generation \cite{louie2020}, recommendations \cite{sanner2023large}, and image generation \cite{chung2023, zhang2023adding}.  Current strategies for providing greater control over a model's operation include targeting model input (e.g., by altering a model's prior probabilities \cite{louie2020}), manipulating underlying representations (e.g., changing the embeddings used for search or image synthesis \cite{cai2019, chung2023}), and/or influencing how model outputs are produced (e.g., constrained decoding techniques \cite{hokamp-liu-2017-lexically}).

One challenge that users may encounter in understanding or controlling an AI's process is that the underlying \textit{representations} may differ from those that a person would use. For example, language models represent words using tokens, which are themselves numerical vectors. Boggust et al. highlight these differences and suggest the value in there being alignment between a person's abstractions and a model's learned abstractions (i.e., abstraction alignment) \cite{boggust2024abstractionalignmentcomparingmodel}.

The AI safety, alignment, and user-centered design communities also emphasize the need for, and value of, providing mechanisms for people to control AI systems \textit{after} they have begun operating. For example, if the system starts to perform unintended or unsafe actions, operators should be able to intervene \cite{russell2016research}. Horvitz and Amershi et al. similarly argue for the value in providing users control in starting and terminating agent processes \cite{horvitz_1999, amershi_2019}. These requirements suggest the value in process \textit{transparency}---the ability to view how the AI is producing an outcome.

Beyond the desired ability to control a model, it is not uncommon for users to want to maintain a sense of \textit{agency} over the outcomes produced (e.g., for artistic endeavors) \cite{chang2023}. This desire has clear implications for alignment, as it means an aligned AI for these use cases is an AI that strikes the right balance between \textit{automation} and \textit{augmentation} \cite{engelbart1962} of end-users. For example, in the example of writing a TODO app, the user may wish to write some of the code themselves to learn how to develop software.

In this paper, we introduce the concept of \textit{process alignment}, which brings together and builds upon these disparate bodies of related work. Process alignment highlights the need for end-users to have transparency, understanding, and control over the manner in which AI systems accomplish tasks. %

\subsubsection{Alignment and Evaluation (Step 3)}
In the final step of the interactive cycle, users must assess the output to determine whether it meets their needs, and to determine what to do next (if anything). Evaluation of the system output can be further divided into two problems: 1) \textit{verifying} the AI's output correctly and completely fulfills the user's intent, and 2) actually \textit{understanding} the AI's output \cite{kim2022, irving2018ai}. In the AI alignment literature, human evaluation has also been considered from the point of view of ensuring \textit{raters} can effectively evaluate the AI's output as part of the model training process (i.e., to provide feedback to improve a reward function used to train the model) \cite{leike2018scalable}. %

For some tasks, verification can be straightforward. For example, it may be relatively easy for a user to assess whether an image produced by a generative model meets their needs. However, in other cases, verification may not be possible or may require extensive work. For example, code synthesis is now a common task for generative text models, but the correctness of arbitrary code is a well-known example in the class of undecidable problems~\cite{turing1936a}. Even when the output of an AI can be proven to be correct, it can still be extremely difficult to do so. For example, despite four years of work by a panel of twelve referees appointed by the Annals of Mathematics, the panel concluded that they could not certify the correctness of a machine-assisted proof of the Kepler conjecture~\cite{hales2005kepler}. %

Thorough \textit{understanding} of a solution can also be challenging \cite{bostrom_2014, morris2023scientists, leike2018scalable, kim2022, irving2018ai}. As an example, Go experts were initially surprised by AlphaGo's Move 37. However, after detailed analysis, it was found to be an excellent move~\cite{menick2016}. Indeed, fully understanding the output of increasingly capable future models that %
can perform better than typical, expert, or even all humans at particular tasks may be theoretically impossible \cite{morris2024levelsagioperationalizingprogress}.

An additional factor that can make evaluation challenging is the large volume of output a generative model can produce in a relatively short period of time. This issue has been observed in music generation \cite{louie2020} and other generative AI systems \cite{buschek2021potential}, with users sometimes feeling overwhelmed with the amount of information produced.

One dimension often considered with respect to model output is explainability, or information about \textit{why} the system produced the output that it did. Amershi et al. and the People + AI Guidebook both advocate for explainability mechanisms in interactive systems to help improve the user experience \cite{amershi_2019, pairguidebook}. Explainability may be helpful for supporting evaluation of model outputs, but understanding \textit{why} the system produced the output is a slightly different problem than helping users verify and understand the output itself. %

In this work, we introduce the concept of \textit{evaluation alignment}, a construct that bridges and builds on this prior work on verification and understanding, %
and offers a novel framework for both analyzing existing systems and innovating future evaluation-support interfaces for end-users of AI-powered systems.

\section{Dimensions of User-Centered Alignment}
Building on the basic three step interaction cycle and user needs detailed above, we describe corresponding user-focused alignment objectives to consider in the design of interactive AI: specification alignment, process alignment, and evaluation alignment. %

\subsection{Specification Alignment}
Starting with the first step of the interaction cycle (input), we define \textit{specification alignment} as the process of the user 1) providing (or the system inferring) a specification of the desired outcome, 2) confirming (i.e., validating) that the system's interpretation of the desired outcome matches their own, and 3) making necessary refinements to that interpretation until sufficient alignment has been reached. In short, the goal is to reach a state where the user is reasonably certain the system has correctly interpreted the user's intent of \textit{what} they want.

Specification alignment focuses on interpretation of intent; how well it actually executes on achieving the desired outcome is a separate issue. For example, a model that can generate photorealistic images may still have low specification alignment if it omits requested details. Conversely, an image synthesis model may include everything requested by the user, but produce an image lacking the desired quality.

As Shen et al. have observed, alignment can also be a bi-directional process \cite{shen2024bidirectionalhumanaialignmentsystematic}. For example, interfaces or interactions that support a user in clarifying their intent (such as Horvitz's suggested human-machine dialogues \cite{horvitz_1999}), help align the AI to the user, whereas interfaces or interactions that elucidate the affordances (or limitations) of a system can help align the user's mental model of the AI and its capabilities. As an example, a user may request a code synthesis AI to produce code for a new programming language it hasn't been trained on. In this case, the AI can alert the user that it is unable to produce code for the target language, thus aligning the user to its capabilities.

\subsubsection{When Specification Alignment Can Occur}
Specification alignment can be achieved \textit{prospectively} (by providing an interpretation of the user's request before any actions are taken), \textit{in tandem} with actions being taken (i.e., a specification is provided while the AI performs actions), or \textit{retrospectively} (a specification of what was done is provided after actions were taken). Providing a specification in parallel with, or after, taking actions may seem counter to the intent of attaining alignment, but taking actions before achieving alignment can lead to efficiencies in instances where the AI correctly interprets the user's intent, or when the AI's output is close enough that editing the final result is at least as efficient as editing the specification (e.g., as can happen in code synthesis \cite{liu2023}). However, there are also circumstances where taking action before specification alignment is achieved is problematic. For example, if the actions an AI can take are costly, have safety implications, or can't be undone, the interaction should be designed so the user can prospectively verify specification alignment.

The goal of specification alignment shares some commonalities with the user-centered ``feedforward'' design principle \cite{vermeulen2013}. A feedforward interface design is an interface that effectively communicates what will happen if a user performs a specific action, before they actually take that action. In a similar spirit, specification alignment seeks to reduce the uncertainty about what an AI will do.

\subsubsection{Specification Refinement and Clarification}
It is possible that users provide under-specified and/or ambiguous specifications to the system (for example, the request ``Write a TODO app'' is under-specified). Thus, specification alignment mechanisms can not only include ways to refine the AI's interpretation of the goal, but also include mechanisms for the AI to request additional information or clarifications from the user (Horvitz's notion of a ``dialogue'' with the user when there is uncertainty \cite{horvitz_1999}). Recent experiments with LLM-based agents provide examples of this concept, with agents requesting clarifications from users before executing the user's request \cite{osika2023}. However, as discussed above, refinement and clarification could happen after partially or fully executing the user's request. For example, a code-generating AI might produce one or more code completions, %
then ask the user which one is closest to their intent before it proceeds.

\subsection{Process Alignment}
In the second step of the interaction cycle (system execution), we define \textit{process alignment} as the ability for the user to directly control and/or influence the actual process by which the AI produces the intended outcome. Whereas specification alignment is focused on reaching agreement on \textit{what} the desired outcome is, process alignment is about reaching agreement on \textit{how} to achieve that outcome. As an example of a process alignment mechanism, the PromptPaint system enables users to control stages of a text-to-image model's generation process by swapping prompts in and out throughout the image generation process \cite{chung2023}. Other examples of process alignment include specifying which corpora a model can reference when generating output (e.g., for code synthesis, only referencing certain corpora when generating code); specifying which models, APIs, or resources to use (e.g., some resources may be more costly than others); or whether computation should be done locally or in the cloud. Importantly, the ability to control a process implies the value in being able to observe, audit, and/or understand the AI's process.

\subsubsection{Post-Process Alignment and Surrogate Processes}
Like specification alignment, process alignment may be achieved before, during, or after the system has produced an output. While it may again seem counterintuitive to align on process \textit{after} the system has already acted, there are several ways post-process alignment can be useful. 

One motivation for aligning on process \textit{after} an action has already been performed is that developing usable mechanisms to reliably control a model may be challenging to implement. In these cases, it may be possible to enhance process alignment by reverse-engineering a simplified, but controllable, representation of the AI's actual process. More specifically, it may be possible for a capable AI to produce an outcome, then reverse engineer a more user-controllable way to produce that outcome (or an outcome substantially similar). When an alternative, controllable formulation of a process is reverse-engineered or made available, we call it a \textit{surrogate process}\footnote{Our use of the term ``surrogate process'' derives from the concept of a \textit{surrogate model}, which refers to a simplified version of a complex model. For example, techniques such as LIME \cite{ribeiro2016} create simplified versions of more complex models to assist in model interpretability.}.  Liu et al.'s system \cite{liu2023} for synthesizing code provides an example of this strategy: After the model generates code from a natural language request, the system reverse-engineers a pseudocode representation of the synthesized code that the user can edit. The edited pseudocode is then used as input to the system to regenerate the actual code. In this case, the pseudocode is a \textit{surrogate} for the actual code: it provides a usable and accessible interface for understanding and controlling the code synthesis process. %

Another motivation for post-output process alignment derives from cases where the user does not have enough information to align on the process up front. As an example, consider a coding assistant that can make use of two models: a fast, but less capable model, and a slower, but more capable model. Depending on the user's needs, they may find that the faster model is sufficient, and not make use of the slower model, or, conversely, feel like they'd rather wait for the slower model given its greater capabilities.

\subsection{Evaluation Alignment}
In the last step of the interaction cycle, the user assesses the output of the AI with respect to their goals. For example, a code synthesis AI may translate the request ``Write a TODO app'' into dozens or hundreds of lines of code, which the user must then assess. \textit{Evaluation alignment} is the process of assisting the user in efficiently and effectively assessing the system's output with respect to their goals.

The depth of evaluation required can vary, so we further differentiate between two levels of evaluation alignment: \textit{verification support}, which focuses on the ability for users to \textit{verify} that the output meets their objectives, and \textit{comprehension support}, which focuses on the ability to \textit{understand} the output. This latter requirement is a much stronger requirement than verification, but an important one to aspire to for systems that rival or exceed users' capabilities or expertise.

The concept of evaluation alignment is an explicit recognition that a successful completion of a human-AI interaction involves the user validating the AI's output. Taking this idea further, evaluation alignment suggests the value in the AI aligning in a personalized manner to a specific user's needs and capabilities. For example, a novice software developer may require greater assistance in understanding a code synthesis model's output compared to an expert software developer. Depending on the coding knowledge of the end-user in our code synthesis example, evaluation alignment support could include one or more of the following:

\begin{enumerate}
    \item Commenting the code.
    \item Creating a natural language explanation of the code produced.
    \item Creating an architectural diagram of the code.
    \item Executing the code in a sandbox, showing the result of code execution.
\end{enumerate}

\section{Case Studies}
In this section, we review a set of existing interactive AI systems to demonstrate how our user-centered dimensions of AI alignment can be used to both describe interfaces and suggest opportunities for further design exploration and/or research. In each system, it is useful to note how the presence of alignment mechanisms provides an enhanced user experience compared to systems that lack these mechanisms.

\subsection{Image Generation: Midjourney and PromptPaint}

For our first case studies, we describe the AI alignment mechanisms for two image generation systems. %

\subsubsection{Midjourney}
Midjourney is a commercial system for producing images using text-to-image models \cite{midjourney}. Its primary interface is text-based: Users enter a prompt describing the intended image and multiple images are produced. Users can influence the image generation process through parameters such as ``chaos,'' which affects the diversity of outputs generated.

Once a set of images has been generated, a user can choose one of the images and invoke an operation to produce variations of that image. The documentation indicates that this operation ``maintains the general style and composition of the selected image.''

To help learn how to craft an effective prompt, users can upload an image and use the ``describe'' command to request four separate text-based prompts that describe the image. There is also a ``shorten'' command to highlight which parts of the prompt were most influential in producing the intended outcome.

Within the alignment dimensions introduced above, the text prompt constitutes the \textit{specification} to the system. In this interface, \textit{specification alignment} is largely supported through post hoc operations such as the ``variations'' operation and the ``shorten'' command: The ``variations'' operation provides a coarse means for users to refine their specification by requesting more images like a chosen image. The ``shorten'' command doesn't provide a means to directly change a specification, but it does provide information that may be helpful in revising the prompt (i.e., the specification). In a similar vein, the ``describe'' command provides an interesting way to deal with the problem of learning how to formulate one's input. This mechanism could be considered an instance of \textit{bi-directional alignment} as it is teaching the user about how to more effectively interact with the AI.

For \textit{process alignment}, parameters like ``chaos'' enable a degree of process control, but these mechanisms are not interactive and no support is provided to understand how a particular value will impact the generated images (i.e., there is no meaningful ``feedforward'' information in the interface design). %

In terms of \textit{evaluation alignment}, images can be quickly reviewed by people, lessening the need for explicit mechanisms to support evaluation. However, one could imagine evaluation mechanisms that show links between specific words in the text inputs and the corresponding regions of the image for those words, to help people verify that objects specified in the prompt are also present in the rendered image.

\subsubsection{PromptPaint}

PromptPaint is a research system that introduces a number of mechanisms to grant the user greater control over text-to-image AI \cite{chung2023}. Like Midjourney, the user provides a text-based prompt to generate an image. However, the user can also specify \textit{multiple} prompts and 1) ``blend'' or interpolate between the prompts using a paint palette metaphor, or 2) request that the system swap input prompts at different points of image generation (for example, prompt A could be used for the first third of the image generation process, and prompt B used for the remainder of the image generation process). The ability to blend prompts using a paint palette metaphor provides a form of \textit{specification alignment}, while the ability to swap prompts during image generation offers a form of \textit{process alignment}.

Like Midjourney, there are opportunities to consider how to support users in evaluating the model output. As an example for this system, a user may wish to produce an image that is a blend of two different painting styles, and want to verify that the end-result captures the stylistic characteristics of each. In a case like this, the AI could produce the requested image, but also produce two additional images from each (unblended) prompt separately. These outputs would help the user visually verify that the image is a blend of the two prompts, but could also aid in debugging in cases where the blended result is not what was expected. In particular, the outputs from the unblended prompts would provide some indication of the model's underlying capabilities; if it is unable to produce a suitable output for an unblended prompt, one would not expect it to be able to produce the desired outcome that is a blending of the two painting styles.

\subsection{Coding: CoPilot and Liu et al.'s Spreadsheet Code Synthesis}

We now consider AI alignment mechanisms for two code synthesis systems.%

\subsubsection{CoPilot}

CoPilot is a commercial product that synthesizes code suggestions from surrounding comments and code \cite{copilot}. %
Code suggestions are automatically generated as users interact with the editor, and users can cycle through code suggestions to choose the suggestion closest to their needs.

In this system, the process of providing a specification can be considered to be both explicit and implicit (i.e., inferred). Within a code editor, users can provide an \textit{explicit} specification by writing comments, documentation, and/or pseudocode that describes the desired code. Implicit specifications are inferred by the model automatically generating suggestions as the user writes and edits code.

CoPilot offers the ability to block suggestions matching public code, which provides some degree of control over its code generation process (\textit{process alignment}). Basic forms of \textit{evaluation alignment} can be achieved if the system comments the code that it synthesizes or if it produces unit tests.

\subsubsection{Liu et al.'s Spreadsheet Code Synthesis}

Liu et al.'s natural language code synthesis system transforms natural language requests into equivalent data processing code \cite{liu2023}. %
However, in contrast to systems like CoPilot, their system does not expose the generated code to the user. Instead, it generates code, then deterministically transforms that code into natural language pseudocode. Users can directly edit this pseudocode then resubmit the pseudocode to regenerate actual code. %

The automatically generated pseudocode can be seen as helping the user attain both \textit{specification} and \textit{process alignment}: The pseudocode reflects the system's interpretation of the user's request, but also the process it will follow to achieve the user's stated goal (i.e., it will translate the given pseudocode to actual code). %
This strategy of echoing back an interpretation of the user's input is similar in spirit to Midjourney's ``describe'' and ``shorten'' commands; in both systems, these echoed specifications can help users learn effective strategies for prompting the models. This reverse-engineering of the final output (i.e., transforming the actual code back into pseudocode) also represents a \textit{surrogate process}---a more accessible representation of the system's actual, underlying process. Finally, because the pesudocode helps users understand the model's output, it also provides \textit{evaluation alignment}.

In a user study conducted by Liu et al. \cite{liu2023}, participants found the pseudocode to be useful to understand what the actual code was doing, and also helpful for debugging. These findings suggest the value of alignment mechanisms in the design of interactive AI systems. %

\section{Discussion}
In this section, we discuss how these alignment dimensions relate to Norman's original Gulfs of Execution and Evaluation; challenges and opportunities for making use of these alignment dimensions; and promising areas for future work.

\subsection{Revisiting Norman's Gulfs}

For decades, Norman's Gulfs of Execution and Evaluation \cite{norman_2002} have been touchstones for guiding the design of interactive systems. With any design, one can ask, ``How can the interactive system help users more efficiently and effectively translate their high-level goals to specific operations?'' (i.e., bridge the Gulf of Execution) and, ``How can the system help users assess the current system state to guide its usage?'' (i.e., bridge the Gulf of Evaluation). These questions highlight fundamental needs any interactive system must address to be useful, usable, and perhaps even delightful.

With the arrival of large language models and generative AI, interactive computing is undergoing a significant shift in \textit{how} people can interact with computational systems (e.g., see Jakob Nielsen's argument referenced earlier \cite{nielsen2023}). As mentioned, interaction with \textit{traditional} computing systems can be characterized as a process in which the user translates high-level goals into specific operations to invoke. In contrast, modern AI enables the user to instead describe the desired outcome, leaving the system to take care of constructing that outcome. While this interaction paradigm enabled by new models is still in its early stages and not without its limitations, it has already proven useful in a number of contexts (perhaps most notably in software development and chatbots). As models continue to improve and become more widespread, it is reasonable to assume that this AI-based interaction paradigm will also become more pervasive, and increasingly \textit{complement} existing interaction paradigms (much as mobile applications grew to complement, but not replace, desktop applications). This likely growth suggests the value in reexamining the users' fundamental needs when using interactive AI systems.

Part of the lasting value of Norman's gulfs is arguably their simplicity and their grounding in the basic interaction cycle. In this paper, we have revisited this interaction cycle in light of the new interaction capabilities afforded by modern AI to define three forms of human-AI alignment: specification, process, and evaluation alignment. Just as Norman's gulfs provide descriptive, evaluative, and prescriptive value, our interactive alignment dimensions provide lenses that one can use to describe, analyze, and explore the design of interactive AI applications. For example, one can ask, ``How does the system help the user specify and/or clarify the desired outcome? What controls and visibility does it provide over \textit{how} the AI will produce an outcome? How does it assist the user in validating and understanding the AI's output?''

In the subsections that follow, we further compare and contrast Norman's original gulfs with the alignment dimensions defined in this paper, and show how these alignment dimensions reveal additional nuance regarding the gulfs an interactive AI design must bridge.

\begin{figure*}[h]
    \centering
    \includegraphics[width=0.75\textwidth]{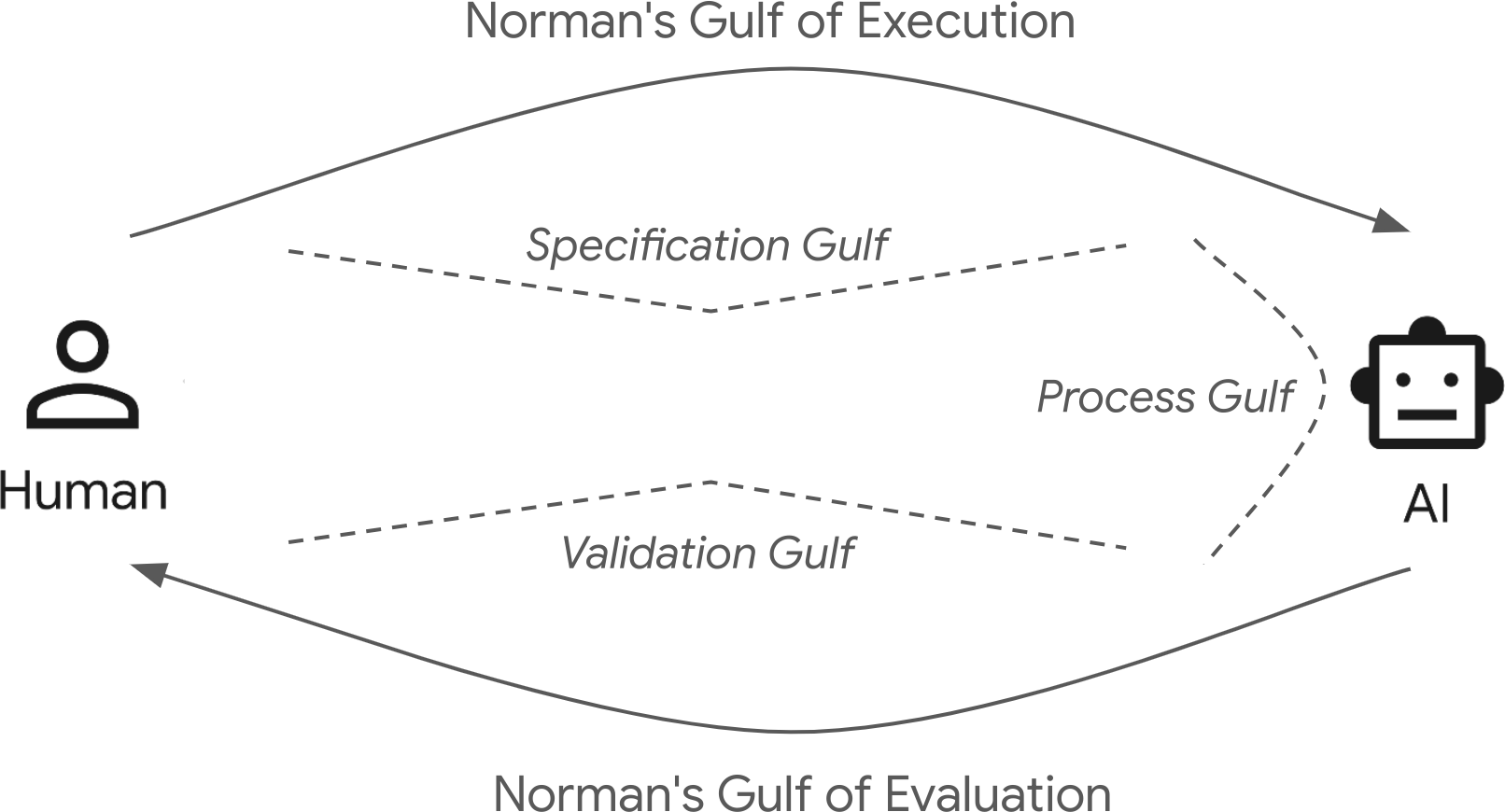}
    \caption{\textbf{Norman's gulfs and the Specification, Process, and Validation Gulfs}. The concept of specification alignment most closely relates to Norman's Gulf of Execution, and implies a Specification Gulf the user must bridge. Process alignment implies a Process Gulf, which spans both the Gulfs of Execution and Evaluation: The user may need to control the AI's process, but may also find it useful to audit the process to better understand the output produced (akin to ``checking one's work''). Evaluation alignment is most closely related to the Gulf of Evaluation, and implies a Validation Gulf the user must bridge to verify and/or understand the AI's output.}
    \label{fig:alignment_gulfs}
\end{figure*}

\subsubsection{Specification Alignment → Specification Gulf}
Specification alignment captures the dual processes of the user correctly and completely communicating their intent to the AI, and the AI effectively communicating its capabilities and limitations to the user. These alignment needs imply a \textit{Specification Gulf} an interactive AI system must bridge in its design.

The Specification Gulf clearly relates to Norman's Gulf of Execution, as both are concerned with the user providing input to the system (Figure \ref{fig:alignment_gulfs}). However, the Specification Gulf highlights a distinguishing characteristic of modern interactive AI systems in contrast with traditional systems: The user describes the desired outcome, instead of providing input to initiate a specific operation.

\subsubsection{Process Alignment → Process Gulf}
When the user needs to understand, modify, or more directly control the AI's process, they must explicitly bridge the \textit{Process Gulf}, which represents 1) the differences between how a human and the AI achieve an outcome, and 2) the different representations employed by each party (see again the abstraction alignment work discussed earlier for a related discussion \cite{boggust2024abstractionalignmentcomparingmodel}). For example, a diffusion model for image generation incrementally transforms an image of pure, statistical noise into a coherent image \cite{saharia2022photorealistic}, an image creation process foreign to most people. 

The Process Gulf can be exacerbated by the fact that understanding how modern AIs function is itself an area of active research.
For example, the area of mechanistic interpretability attempts to understand how an AI-trained algorithm solves a problem. In this research area, it has been observed that even a seemingly simple problem, such as explaining how a one layer neural network learns to do modular arithmetic, requires significant effort to understand~\cite{grokkingexplorable2023}.  Fortunately, because AI systems can be understood at many levels of abstraction, understanding low-level implementation details (e.g., the exact algorithm learned by an AI) is often not needed to provide some form of meaningful control to end-users. Additionally, strategies such as surrogate processes (previously described) can help bridge the Process Gulf.

In relation to Norman's original Gulfs of Execution and Evaluation, the Process Gulf is most closely related to the Gulf of Execution, but also spans the Gulf of Evaluation (Figure \ref{fig:alignment_gulfs}). Like the Gulf of Execution, the Process Gulf describes the challenges a user faces in determining how to effectively control an underlying computational process. However, \textit{how} the AI produces an output can also be useful in understanding that output once it has been produced. For example, if an AI produces an output through a number of sequential reasoning steps, showing those reasoning steps can be useful in assessing that output.

Our observations that the Gulf of Execution for modern AI systems comprises both a Specification Gulf and a Process Gulf (Figure \ref{fig:alignment_gulfs}), and that the Process Gulf also has relevance in helping the user understand output, can assist designers of AI models and systems in better addressing user needs.%

\subsubsection{Evaluation Alignment → Validation Gulf}
While Norman's Gulf of Evaluation focuses on assessing the current system state, the \textit{Validation Gulf} refers to the challenge a user faces in verifying and understanding an AI's output. Verifying and/or understanding an AI's output can be challenging when the AI's output exceeds the user's expertise, or when the form of the output is not optimized for the user's perceptual and/or cognitive capabilities. As an example of the latter, an AI may produce hundreds of lines of code in a matter of seconds. While the user may have the expertise and ability to review the code, there may be additional or alternative outputs an AI could produce to better support evaluation. 

\subsubsection{Strategies for Bridging Alignment Gulfs}
There are at least two high-level strategies for bridging alignment gulfs: through model improvements and through interaction design. As an example, initial large language models (LLMs) supported prompting \cite{brown2020language}, but were not especially good at following instructions. This limitation represents a Specification Gulf. One of the more effective approaches at addressing this gulf from the modeling end was through instruction tuning \cite{ouyang2022training}. In other contexts, the specification alignment has been addressed through interactive mechanisms, such as systems that interactively ask the user clarifying questions about their goals (e.g., \cite{osika2023}).

Breaking down Norman's original two gulfs for traditional systems into three gulfs for modern AI systems also helps us identify an important theoretical concern and related set of research and practical challenges. While Norman's original gulfs were meant to be addressable through appropriate system and interaction design, each of our three gulfs may be theoretically impossible to fully bridge for sufficiently advanced AI systems. For instance, consider the Specification Gulf: It may be impossible for an AI to correctly and fully communicate its capabilities to the end-user, particularly if it is a very general system that might have countless uses or even emergent capabilities (i.e., the ability to perform tasks beyond those the model was initially trained for \cite{morris2024levelsagioperationalizingprogress}). Or, consider the Process Gulf: Currently, modern neural networks' operations are not interpretable, even to machine learning (ML) experts; without enormous advances in mechanistic interpretability research, the Process Gulf may also not be fully bridgeable. Finally, consider the Validation Gulf: As frontier models progress in capabilities (possibly up to future theorized systems that could be considered artificial general intelligence or superintelligent \cite{morris2024levelsagioperationalizingprogress}), the Validation Gulf may become theoretically impossible to completely bridge for many user-task combinations. This framing opens up an important class of research questions (namely, to find ways to fully bridge these seemingly impossible gulfs for AGI models and their precursors) and practical challenges (specifically, to develop interactions and model interventions that reduce each of these three gulfs to the greatest extent possible). %

\subsection{Appropriate Levels of Alignment: Helping, Not Hindering Task Progression}
The notion of achieving AI alignment suggests a process that needs to be completed before any action can be taken. However, in human-AI interactions, it can be cumbersome or impractical to reach alignment on \textit{everything} before an action is taken if the costs of performing a (potentially incorrect) action are negligible. For example, for code synthesis, rather than requiring a detailed specification of what code to produce, it may be more efficient to simply generate the code and make refinements to 1) the initial request, 2) the resulting code, or 3) some representation of the generated code (e.g., pseudocode). CoPilot employs this strategy, with an interface design that allows the user to dismiss or ignore irrelevant or unneeded suggestions.

With these points in mind, the alignment objectives described in this paper should not be interpreted as an argument for reaching complete alignment before the AI takes any action. As mentioned above, this may not be practical or provide the optimal user experience. Instead, what these issues highlight is a design challenge for creating \textit{appropriately aligned} AI systems: How can we design interfaces that enable users to reach \textit{appropriate levels of alignment} with the AI, without negatively affecting the user experience (and without the potential for negative side effects)? CoPilot and Liu et al.'s code synthesis system both provide examples of system designs that strike a balance between moving forward and providing facilities to make corrections as necessary \cite{liu2023}.

\subsection{The Challenge of Supporting Evaluation}

As AI capabilities increase, the need for users to be able to verify and understand the AI's output also increases. This need has been the focus of significant research in the AI alignment community in the context of training aligned AI. For example, Irving et al. consider how self-debate could be used to train AI that produces true, useful information \cite{irving2018ai}. With the similar goal of creating aligned AI, Leike et al. consider how to evaluate AI in problem domains where it is difficult for people to evaluate the correctness of a solution (e.g., because of complexity, lack of expertise, etc.) \cite{leike2018scalable}. Their proposed strategy of ``recursive reward modeling'' includes the notion of training agents that help people evaluate individual facets of the AI's output, rather than asking people to evaluate the entire output all at once (e.g., asking a person to evaluate summaries of character arcs in a novel, as opposed to asking them to evaluate the entire novel). To improve the quality of model outputs when performing chain-of-thought reasoning \cite{wei2023chainofthought}, Lightman et al. showed that process supervision (i.e., improving the model's reasoning process) can lead to better output \cite{lightman2023lets}.

While this prior work is largely concerned with optimizing AI performance and alignment during AI training, the methods developed have direct applicability to the design of mechanisms intended to support evaluation of AI output in interactive AI systems. For example, Irving et al.'s self-debate strategy and Leike et al.'s recursive reward modeling will both yield AI that produce outputs that are easier for people to evaluate when interactively using AI. Similarly, an improved and more reliable chain-of-thought reasoning process can be exposed to the user to help them evaluate the outcome by seeing how it was derived, step-by-step.

This prior work provides a useful foundation and inspiration for evaluation support mechanisms, but there are significant opportunities for additional research into how to support people in reliably and efficiently verifying and understanding AI outputs.

\subsection{Alignment with Multiple Parties: CSCW and AI Alignment}
Our discussions of achieving alignment have assumed a 1:1 dynamic with a single user interacting with a single AI. However, it is also useful to consider alignment for interactions that include multiple entities, as this can significantly alter interaction requirements (e.g., see Suh et al.'s study examining how collaborations can be affected when an AI is introduced into a music creation task involving two people \cite{suh2021}). In these contexts, the same alignment goals still exist, with the added dimension of attaining alignment across all parties involved.

As an example, consider a scenario of a team working on a larger design problem. In this scenario, one could imagine the team jointly developing a specification in real-time with the AI. This joint development could happen serially, with a single person distilling the group's work into a single specification, but there are interesting possibilities where the AI facilitates the process of the group coming into consensus \textit{in parallel}. In this instance, specification alignment is happening not only between users and the AI, but also \textit{amongst} users. Once the AI produces an output, there are also interesting opportunities for the AI to assist each team member individually in understanding the output (e.g., providing representations and/or capabilities matching each individual's expertise).

More generally, one can think about interactive alignment in computer-supported collaborative work  (CSCW) contexts: What unique interface and interaction needs are introduced when AI is used individually or collectively by a group working toward a common goal?  This interactive perspective on multi-user AI alignment differs from the focus on \textit{pluralistic alignment} in the ML community \cite{Gabriel_2020}, which is primarily concerned with the (important, but potentially intractable) challenge of creating a model that incorporates the moral values of diverse individuals and cultures. In contrast, a consideration of interactive alignment suggests a need to consider how to design interfaces that best support specification, process, and evaluation alignment for groups of users jointly operating an AI. Model advances supporting pluralistic value alignment and interface advances supporting multi-party interactive alignment are both valuable avenues for future research, and may complement each other in many situations (i.e., it may be easier to develop 
multi-user interfaces with a model that is pluralistically aligned).

\section{Conclusion}
This paper has joined related literature in the AI alignment and HCI communities to synthesize three user-centered objectives for alignment: specification alignment (aligning on \textit{what} outcome to produce), process alignment (aligning on \textit{how} to produce that outcome), and evaluation alignment (helping the user verify and understand what was produced). We showed how these alignment perspectives could be used to describe the interfaces of existing interactive AIs, while also suggesting ways they could be enhanced. We also examined how these alignment dimensions relate to Norman's original Gulfs of Execution and Evaluation, and reflected on how interaction with modern AI systems might be more precisely represented by a three-gulf model, corresponding to our alignment dimensions.

Looking ahead, there are opportunities to build on this foundation to identify how user-centered, interactive alignment can be considered in a broader range of scenarios, such as those involving multiple simultaneous users and/or AIs.

\begin{acks}
We thank Iason Gabriel, Geoffrey Irving, Michael Liu, Carrie Cai, and Adam Pearce for helpful discussions and feedback. We also thank Aaron Donsbach for producing our figures.
\end{acks}

\bibliographystyle{ACM-Reference-Format}
\bibliography{refs}

\appendix

\end{document}